\documentclass[aps,prb,twocolumn,showpacs,amsmath,amssymb,groupedaddress,floatfix,longbibliography]{revtex4-1}
\usepackage{graphicx,color}
\usepackage{amsmath,amssymb,bm}
\usepackage{epstopdf}
\usepackage{braket}
\setcitestyle{square,numbers}

\usepackage[plainpages=false,pdfpagelabels,colorlinks=true,linkcolor=red,urlcolor=blue,citecolor=blue,pdftitle={Title},pdfauthor={},pdfdisplaydoctitle=true,pdfduplex=DuplexFlipLongEdge]{hyperref}

\begin{document}
\title{Density-matrix renormalization group method for the conductance of one-dimensional correlated systems 
using the Kubo formula}

\author{Jan-Moritz Bischoff}
\email[E-mail: ]{jan.bischoff@itp.uni-hannover.de}
\author{Eric Jeckelmann}
\affiliation{Leibniz Universit\"{a}t Hannover, Institut f\"{u}r Theoretische Physik, Appelstra\ss e 2, D-30167 
Hannover, Germany}

\date{\today}

\begin{abstract}
We improve the density-matrix renormalization group (DMRG) evaluation of the Kubo formula for the zero-temperature linear conductance
of one-dimensional correlated systems.
The dynamical DMRG is used to compute the linear response of a finite system to an applied AC source-drain voltage,
then the low-frequency finite-system response is extrapolated to the thermodynamic limit to
obtain the DC conductance of an infinite system.
The method is demonstrated on the one-dimensional spinless fermion model at half filling.
Our method is able to replicate several predictions of the Luttinger liquid theory such as the renormalization of the 
conductance in an homogeneous conductor, the universal effects of a single barrier, and the resonant tunneling through a double barrier.
\end{abstract}


\maketitle

\section{Introduction}

Electronic systems exhibit a number of interesting properties when they are confined to reduced spatial dimensions. 
In particular, the transport properties of (quasi-)one-dimensional correlated electron systems such as quantum wires 
have been
extensively studied during the last two decades~\cite{baeriswyl04,giamarchi03,kawa07}.
They differ vastly from the well understood dynamical properties of a three-dimensional metal.
The theory of Luttinger liquids describes the low-energy properties
of one-dimensional correlated conductors~\cite{giamarchi03}.
Electronic  Luttinger liquids are
believed to be realized in semiconductor quantum wires~\cite{tarucha95a}, carbon nanotubes~\cite{bockrath99a},
and atomic wires deposited on semiconducting substrates~\cite{blumenstein11a,ohtsubo15a}.
Beyond the generic Luttinger liquid paradigm, however,
we only have a fragmentary understanding of quantum transport in
one-dimensional systems because we lack strong versatile methods for these problems.

The density-matrix renormalization group (DMRG) method is the most powerful numerical method
for computing the properties of one-dimensional correlated lattice models~\cite{whit92b,whit93a,scho11,jeck08a}.
Various approaches have been developed to compute the transport properties using DMRG.
In particular, time-dependent DMRG simulations of systems driven out of equilibrium
have proven to be a useful tool for this purpose.
They have been used successfully to investigate the conductance of
small interacting systems coupled to noninteracting leads~\cite{schm04a,al-hassanieh06a,boulat08a,kirino08a,fabian09a,fabian09b,bran10,fabian10a}
and of isolated quantum wires out of equilibrium~\cite{schm04a,kirino10a,einh12}.
In addition, time-dependent  DMRG approaches have been developed to study
the Drude weight of quantum systems at finite temperature~\cite{karrasch13a,karrasch14a,karrasch16a}
as well as non-equilibrium steady states in quantum spin chains using the Lindblad formalism~\cite{prosen09a,znidaric11a}.
However, it remains very difficult to carry out accurate calculations over a long enough period of time to simulate
the DC transport in large systems. Therefore, a DMRG method that computes DC properties such as the conductance of quantum wires directly
is very desirable.

For instance, the static response to twisted boundary conditions~\cite{kohn64}
was used to compute the Drude weight (charge stiffness) of correlated chains~\cite{schm04b,dias06a}
and the conductance through a short interacting region inside a noninteracting ring~\cite{molina03a,meden03a}.
However, this approach is not practical
because of the lower efficiency of DMRG methods for systems with periodic boundary conditions.

A decade ago, Bohr et al.~\cite{Bohr2006} showed that DMRG could be used to evaluate 
the Kubo formula for the linear response to a potential bias~\cite{kubo57a}.
(Similarly, DMRG and Kubo formalism can be combined to compute the Drude weight~\cite{schm04b,shir09}.)
Such DMRG computations of the linear conductance were carried out 
for various interacting systems coupled to noninteracting leads:
short wires~\cite{Bohr2006}, small nanostructures~\cite{bohr07a}, and benzene-like ring structures~\cite{bohr12a}.
Surprisingly, this approach has rarely been used.

In this paper we revisit and improve the DMRG evaluation of the Kubo formula for the zero-temperature linear conductance
of one-dimensional correlated lattice models.
We first show how to compute the linear response of a finite system to an applied AC source-drain voltage using the dynamical
DMRG method~\cite{jeck02a,jeck08b} and then how to extrapolate the low-frequency finite-system response
to the thermodynamic limit to obtain the DC conductance of an infinite chain.
The method is demonstrated on the one-dimensional spinless fermion model at half filling.
We show that our approach is able to replicate several predictions of the Luttinger liquid theory, 
namely the renormalization of the conductance in an homogeneous conductor~\cite{apel82a},
the universal effect of a single barrier (on-site impurity)~\cite{kane92a,Kane1992},
and the surprising resonant tunneling through a double barrier~\cite{Kane1992}.

\section{Model and method}

\subsection{Model}

We consider a one-dimensional lattice model with $M$ sites and open boundary conditions.
The Hamiltonian of the unperturbated system is
\begin{eqnarray}
H & = & -t\sum\limits_{j=2}^{M}
\left ( c^{\dagger}_{j}c^{\phantom{\dagger}}_{j-1}+c^{\dagger}_{j-1}c^{\phantom{\dagger}}_{j} \right ) \nonumber \\
&& + V\sum\limits_{j=2}^{M}\left ( n_{j} -\frac{1}{2} \right ) \left ( n_{j-1} -\frac{1}{2} \right ),
\label{eq:ham}
\end{eqnarray}
where $c^{\dagger}_{j}$ ($c^{\phantom{\dagger}}_{j}$) creates (annihilates) a spinless fermion on site $j\ (=1,\dots,M)$
and $n_{j}=c^{\dagger}_{j}c^{\phantom{\dagger}}_{j}$ is the density operator on the same site.
We focus on the half-filled chain, i.e, the number of fermions is $M/2$ and the system length is even.
In addition, we assume that the hopping term $t > 0$.
This model is exactly solvable using the Bethe Ansatz method~\cite{giamarchi03,sirk12}.
Its excitation spectrum is gapless in the thermodynamic limit for the nearest-neighbor interaction parameter $-2t < V \leq 2t$
and its low-energy properties are then described by the Luttinger liquid theory.
In this work we discuss only the model properties in this Luttinger liquid phase.

\subsection{Conductance \label{sec:conductance}}

\begin{figure}
\includegraphics[width=.99\columnwidth]{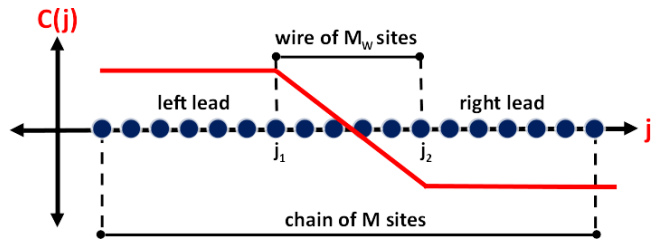}
\caption{
Scheme of the system. A one-dimensional lattice of $M$ sites 
is divided into three segments by an external potential with a profile determined by coefficients $C(j)$
in Eq.~(\ref{eq:potential}): left and right leads where the potential is constant and quantum wire of $M_{\text W}$
sites where the potential decreases linearly.
}
\label{fig:1}
\end{figure}

Following~\cite{kane92a,Kane1992} we apply a spatially constant electric field in a restricted interval of a long wire.
Thus the chain is separated into three segments (see Fig.~\ref{fig:1}) that play the role of a left lead, a quantum wire, and 
a right lead.   
Note that the hopping amplitude $t$ and interaction parameter $V$ remain uniform as defined in the Hamiltonian~\eqref{eq:ham}. 
Therefore, the distinction between leads and wire emerges solely from the applied field.

To generate a current in a finite system we assume that the applied electric field oscillates slowly in time. This 
results in a time-dependent perturbation 
\begin{equation}
\delta H(\tau)= q V_{\text{SD}} f(\tau) \sum\limits_{j=1}^{M}C(j)n_{j},
\label{eq:perturbation}
\end{equation}
where $q$ is the charge carried by one spinless fermion,
$V_{\text{SD}}$ is the potential difference between source and drain (left and right leads), 
$f(\tau)$ is a dimensionless function
of time $\tau$ oscillating between -1 and 1, and the potential profile is given by
\begin{equation}
C(j)=
  \begin{cases} 
      \hfill \hphantom{-} \frac{1}{2}    \hfill & \text{ for $j\leq j_{1}$} \\
      \hfill -\frac{j-j_{1}}{j_{2}-j_{1}}+\frac{1}{2}    \hfill & \text{ for $j_{1} < j < j_{2}$} \\
      \hfill -\frac{1}{2} \hfill & \text{ for $j\geq j_{2}$} .
  \end{cases}
\label{eq:potential}
\end{equation}  
The wire includes the $M_W (=j_2-j_1+1)$ sites with indices $j_1 \leq j \leq j_2$, see Fig.~\ref{fig:1}.
  
The time-dependent perturbation~(\ref{eq:perturbation}) generates a current in the system.
We focus on the current flowing through the wire.
The corresponding current operator is
\begin{equation}
J=\frac{1}{M_{W}-1} \frac{it}{\hbar}  \sum\limits_{j=j_{1}+1}^{j_{2}} 
\left ( c^{\dagger}_{j}c^{\phantom{\dagger}}_{j-1}-c^{\dagger}_{j-1}c^{\phantom{\dagger}}_{j} \right ).
\label{eq:current}
\end{equation}
The frequency-dependent linear conductance is then defined by 
\begin {equation}
G(\omega) = \lim_{V_{SD}\rightarrow 0}  \text{Re} \left \{ \frac{q\langle J \rangle(\omega)}{V_{SD}f(\omega)} \right\}
\label{eq:conductance}
\end{equation}
where $f(\omega)$ and $\langle J \rangle(\omega)$  denote the Fourier transforms of the function $f(\tau)$ in~(\ref{eq:perturbation})
and the expectation value of the current operator~(\ref{eq:current}), respectively.
The DC conductance is the zero-frequency value 
\begin{equation}
G=G(\omega \rightarrow 0).
\label{eq:conductance2}
\end{equation} 

As we consider a spinless fermion wire, the quantum of conductance is 
\begin{equation}
G_0=\frac{q^2}{h}.
\label{eq:quantum}
\end{equation} 
In all our numerical results the energy scale is set by $t=1$, the charge by $q=1$, and $\hbar=1$.
This yields $G_0 = \frac{1}{2\pi}$. Therefore, we show $2\pi G=G/G_0$ in our figures.

\subsection{Kubo formula}

Applying the Kubo formula for the linear response of the model~(\ref{eq:ham}) at zero temperature 
to the perturbation~(\ref{eq:perturbation}) yields 
\begin{equation}
G(\omega) = \lim\limits_{\eta\rightarrow 0^+} \frac{q^2}{\omega} \left [ G_{J,\eta}(\omega) - G_{J,\eta}(-\omega) \right ]
\label{eq:kubo}
\end{equation}
with the imaginary part of the dynamical current-current correlation function
\begin{equation}
G_{J,\eta}(\omega) = \left \langle 0 \left \vert J \frac{\eta}{(E_{0}-H+\hbar \omega)^2+\eta^2} J \right \vert 0 \right \rangle,
\label{eq:correlation}
\end{equation}
where the expectation value is calculated for the ground state of the unperturbated Hamiltonian~(\ref{eq:ham}) with energy $E_0$.
Correlation functions~(\ref{eq:correlation}) can be calculated accurately for fixed frequencies in 
one-dimensional correlated quantum models using the dynamical DMRG method~\cite{jeck02a,jeck08b}.

Bohr et al.~\cite{Bohr2006} calculated the difference in~(\ref{eq:kubo}) analytically and obtained another correlator
[i.e., the derivative of~(\ref{eq:correlation}) with respect to $\omega$], which  they evaluated directly at zero-frequency using DMRG.
Here, we have chosen to compute the correlation function~(\ref{eq:correlation}) for a narrow frequency interval around $\omega=0$ with 
dynamical DMRG 
and then to calculate the difference in~(\ref{eq:kubo}) numerically for $\omega\rightarrow 0$. 
We were able 
to evaluate the Kubo formula for chains with up to $M=2000$ sites using the dynamical DMRG with less than $m=200$ density-matrix eigenstates kept.
For comparison, Bohr et al.~\cite{Bohr2006} reported using up to $m=1200$ density-matrix eigenstates for systems with up to 200 sites only.

\begin{figure}
\includegraphics[width=.8\columnwidth]{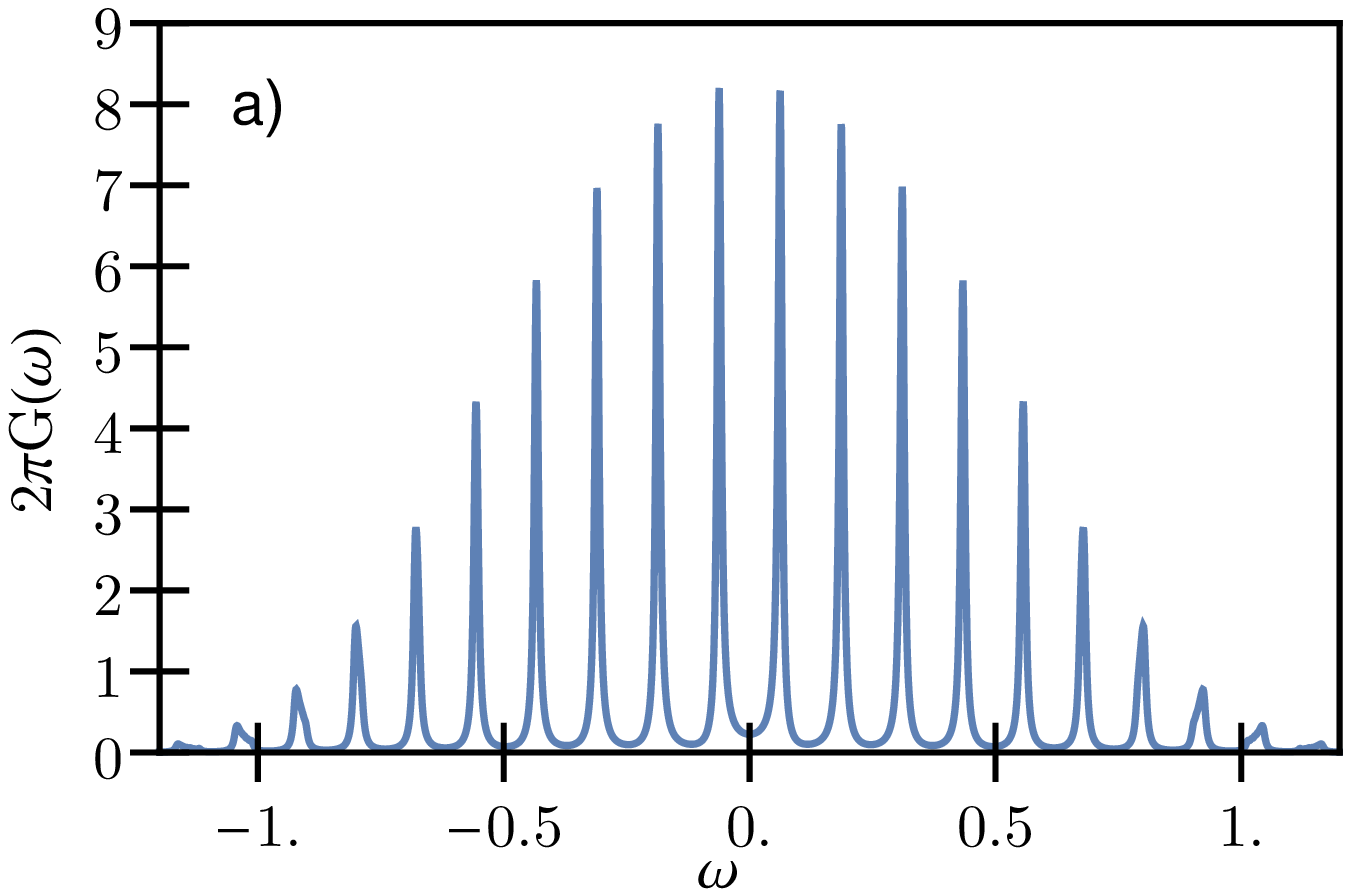}
\includegraphics[width=.8\columnwidth]{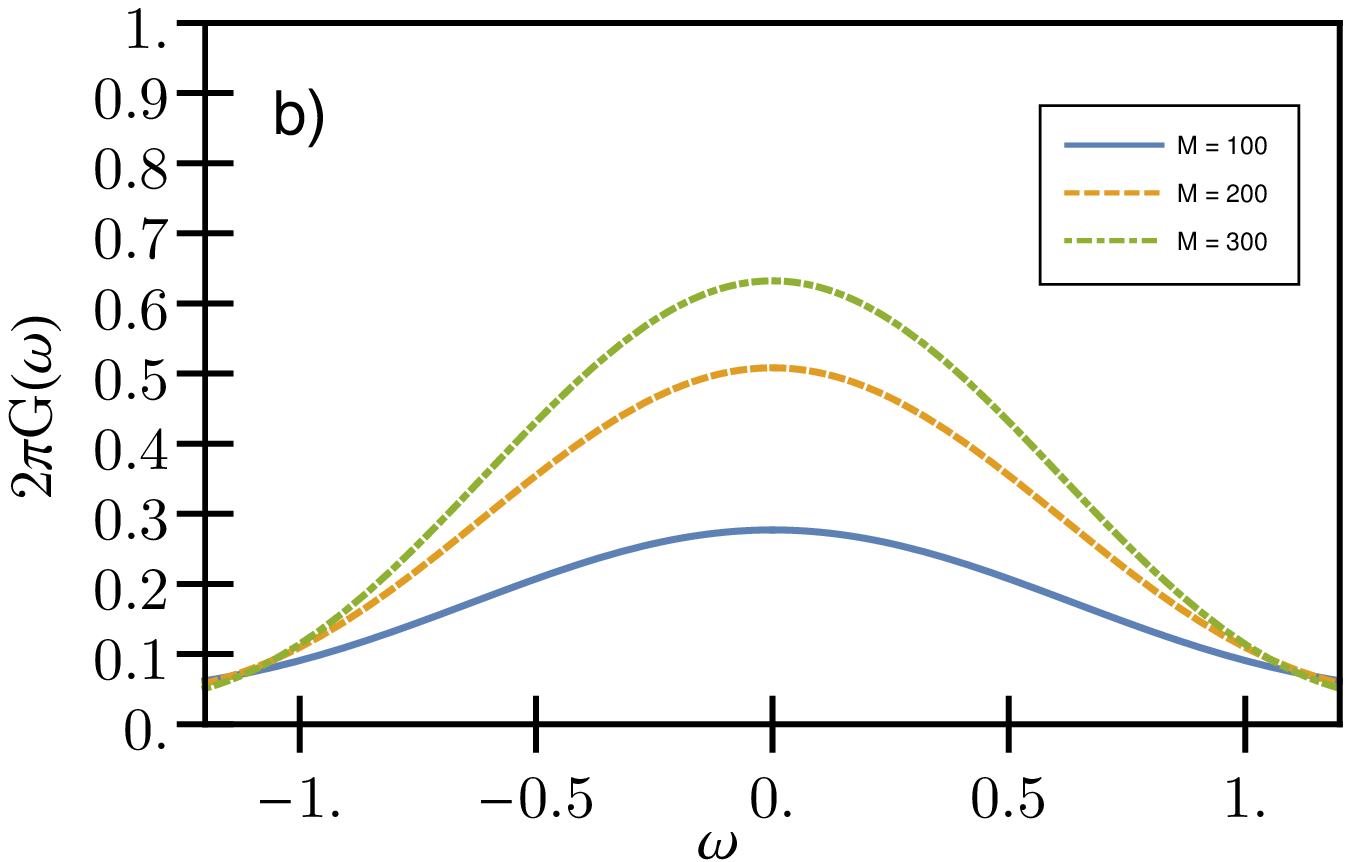}
\caption{
AC conductance $G(\omega)$ of the noninteracting chain for 
(a) $M=100$ with broadening $\eta = 0.48t/M$
and (b) three different system sizes $M$ with broadening $\eta = 48t/M$. 
In all cases $M_{W}=10$. 
}
\label{fig:spectra}
\end{figure}

Obviously, our approach is both simpler and computationally faster 
but one could fear that it is ill-conditioned because of the divergent term $1/\omega$ multiplying the
difference between correlation functions in the Kubo formula~(\ref{eq:kubo}). 
However, this formula corresponds to taking the derivative of~(\ref{eq:correlation})
for $\omega\rightarrow 0$ and we have found that this correlation function is smooth around $\omega=0$ if the limit $\eta\rightarrow 0$ 
is taken properly. 
Problems occur only for finite systems (i.e., taking the limits $\eta, \omega \rightarrow 0$ for a fixed $M$).
Therefore, the real issue is to use the proper finite-size scaling.
As an example, Fig.~\ref{fig:spectra} shows $G(\omega)$ for an noninteracting chain ($V=0$)
for various values of $M$ and $\eta$.
Figure~\ref{fig:spectra}(a) reveals the discrete structure of the spectrum for a too small broadening $\eta$
while Fig.~\ref{fig:spectra}(b) illustrates the smooth spectra $G(\omega)$ that are obtained with large enough broadening $\eta$.

\subsection{Finite-size scaling}

Dynamical DMRG yields numerical results for $G_{J,\eta}(\omega)$ in a system with finite sizes $M$ and $M_{\text{W}}$ at finite broadening $\eta$.
Steady-state transport is ruled out in a finite-length chain with open boundary conditions, however.
For instance, the Drude spectral weight of a one-dimensional metal is shifted to a finite frequency $\sim 1/M$ 
in the optical conductivity spectrum~\cite{fye91a}.
Therefore, taking~(\ref{eq:conductance2}) and~(\ref{eq:kubo}) into account, we have to compute three limits
 \begin{equation}
 G=\lim_{\omega\rightarrow 0} \lim_{\eta \rightarrow 0^+} \lim_{M \rightarrow \infty} 
\frac{q^2}{\omega} \left [ G_{J,\eta}(\omega) - G_{J,\eta}(-\omega) \right ]
 \label{eq:extrap}
 \end{equation}
 from our DMRG data for $G_{J,\eta}(\omega)$.
This is the physically correct order of the limits. The time required to go through the system
$\sim M$ must be larger than the measurement time $\sim 1/\eta$, which must be larger than the period of 
the perturbation $\sim 1/\omega$ in the linear response theory.
Note that changing the limit order can yield wrong results. For instance, 
taking the limit $M$ last always yields $G=0$.  

In agreement with the finite-size-scaling analysis of the optical conductivity presented in Ref.~\cite{jeck02a},
we have found that we can take the first two limits simultaneously using the scaling $\eta M = C$ where
the constant $C$ is large enough to hide the discrete structure of the finite-system spectra.
Then the value of the conductance can be obtained directly at zero-frequency because the finite $\eta$ smoothens 
the spectrum of $G(\omega)$ over a range $\Delta\omega \approx \eta = C/M$ around $\omega=0$. 
For instance, the smoothened spectra can be seen in Fig.~\ref{fig:spectra}(b) for a noninteracting chain
using $\eta M = 48t$. 

Physically, this scaling means that we can simulate DC transport over a finite time scale $\sim \omega^{-1} \sim \eta^{-1} \sim M$
in a finite system of size $M$.
We note $G(M)$ the value of the conductance obtained with this procedure for a fixed system size $M$,
\begin{equation}
G(M) = \left . \frac{q^2}{\omega} \left [ G_{J,C/M}(\omega) - G_{J,C/M}(-\omega) \right ] \right \vert_{\omega=0}.
\end{equation}
Extrapolating these values to $M\rightarrow\infty$ yields the DC-conductance in the thermodynamic limit~(\ref{eq:extrap}).
For all results presented here, we have used $\eta=48t/M$ and chain lengths up to $M=2000$.    

\begin{figure}
\includegraphics[width=.9\columnwidth]{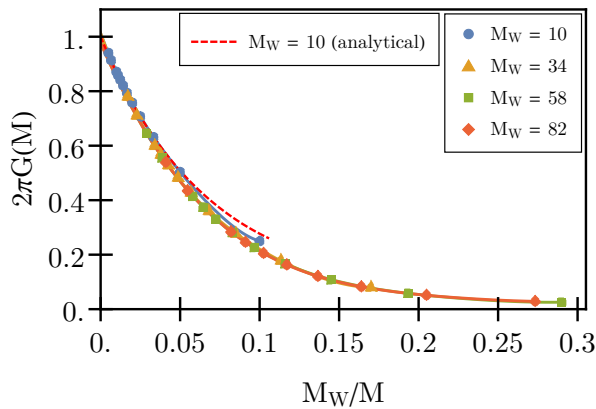}
\caption{
DMRG results for the conductance $G(M)$ of the noninteracting chain as a function of the ratio $M_{\text W}/M$ between system size and wire length
for different wire lengths $M_{\text W}$. 
For all cases $\eta = 48t/M$. Solid lines are polynomial fits. The dashed line shows the exact result for $M_{\text{W}}=10$.
}
\label{fig:scaling}
\end{figure}

For a noninteracting chain [i.e., $V=0$ in the Hamiltonian~(\ref{eq:ham})] we can perform all calculations analytically
and we recover the quantum of conductance~(\ref{eq:quantum}) as expected.
For finite $M$ and $M_{\text{W}}$ we can reduce the expectation value~(\ref{eq:correlation}) to sums
over the single-particle eigenstates, which can be evaluated exactly using simple numerics.
The resulting $G(M)$ is shown in Fig.~\ref{fig:scaling} for $M_{\text{W}}=10$.

With this procedure we obtain the DC-conductance $G$ for a wire of finite size $M_{\text{W}}$. While
there are open problems for finite-size structures that we could investigate with this approach,
we are here interested in long wires that can exhibit the properties of Luttinger liquids.
Therefore, we also have to analyze the finite-size scaling with $M_{\text{W}}$.

Furthermore, a space- and time-dependent electric field $E(k,\omega)$ with wave number $k$ and frequency $\omega$
induces a current $j(k,\omega) = \sigma(k,\omega) E(k,\omega)$ in a one-dimensional
system with linear conductivity $\sigma(k,\omega)$~\cite{schulz95a}.
In an ideal, infinitely long one-dimensional conductor, the conductance $G$ determines
the response to an inhomogeneous ($k\neq 0$) but static ($\omega=0$) field: $\text{Re} \{ \sigma(k,0) \} = 2\pi G \delta(k)$.
In the setup of Fig.~\ref{fig:1} with $M\rightarrow \infty$, the wave number is solely determined by the range
over which the potential varies, i.e. 
$k \sim 1/M_{\text{W}}$, and thus we have to investigate the scaling of $G$ for large enough $M_{\text{W}}$
after taking the limits in~(\ref{eq:extrap}). 
In contrast, calculating the limit $M\rightarrow \infty$ for a fixed ratio $M/M_{\text{W}}$ first and 
then taking the limit $\omega,\eta \rightarrow 0$ yields the Drude weight $D$, which determines
the response to an homogeneous ($k=0$) but time-dependent($\omega \neq 0$) electric field: 
$\text{Re} \{ \sigma(k=0,\omega) \} =2\pi D \delta(\omega)$.  

As a first test, we computed the dynamical correlation function~(\ref{eq:correlation}) with DMRG for the 
noninteracting system [$V=0$ in Eq.~(\ref{eq:ham})]
and then calculated $G(M)$ as described above. The results are shown in Fig.~\ref{fig:scaling}.
Clearly, $G(M)$ converges toward the exact result~(\ref{eq:quantum}) for increasing $M$ and the convergence 
depends mostly on the ratio $M/M_{\text{W}}$ between system size and wire length.
Physically, this just means that the charge reservoirs (both lead parts of the one-dimensional lattice) must
be much larger than the central wire segment to simulate DC transport in the wire.
Figure~\ref{fig:scaling} shows that we can reproduce the exact result using relatively small wires.
This is very convenient because large ratio $M/M_{\text{W}}$ are needed and 
the DMRG computational cost increases rapidly with $M$.

Therefore, we conclude that our approach is accurate enough to determine the DC conductance in the thermodynamic limit.
In the next section, we will illustrate its possibilities for interacting systems using a fixed wire length $M_{\text{W}}=10$,
first for homogeneous systems and then for wires including one or two site impurities.

\section{Results}

\subsection{Homogeneous Luttinger liquid}

According to the theory of Luttinger liquids, the transport properties of an ideal one-dimensional 
conductor can be renormalized by the interaction between charge carriers.
In particular, the DC conductance of an homogeneous one-channel Luttinger liquid is~\cite{apel82a,kane92a,Kane1992}
\begin{equation}
G_{\text{LL}} = K G_0,  
\label{eq:luttinger}
\end{equation}
where $K$ is the so-called Luttinger parameter.
This result is valid for the setup described in Sec.~\ref{sec:conductance}, i.e.
when the interaction parameters are identical in leads and wire.
Its relevance for experiments is still controversial~\cite{kawa07}.
For the half-filled spinless fermion model~(\ref{eq:ham}), the Luttinger parameter
can be calculated exactly
from the Bethe Ansatz solution of the 1D spin-$\frac{1}{2}$ Heisenberg model~\cite{sirk12}
\begin{equation}
K = \frac{\pi}{2}\frac{1}{\pi-\arccos\left(\frac{V}{2t}\right)}.
\label{eq:bethe}
\end{equation}
Therefore, the DC conductance is known exactly for an homogeneous Luttinger liquid.

\begin{figure}
\includegraphics[width=.9\columnwidth]{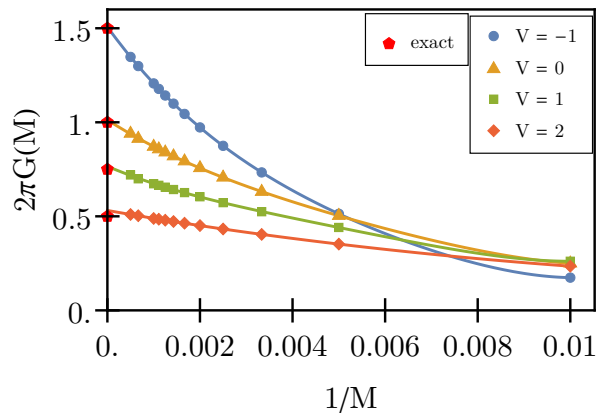}
\caption{
DMRG results for the conductance $G(M)$ as a function of the inverse system size  
for various nearest-neighbor interaction strengths $V$. Solid lines are polynomial fits. 
For all cases $\eta = 48t/M$ and $M_{\text{W}}=10$. 
Pentagons on the vertical axis ($1/M=0$) 
show the exact values predicted by the Luttinger liquid theory~(\ref{eq:luttinger}) combined with the Bethe Ansatz solution~(\ref{eq:bethe}).
}
\label{fig:homogeneous}
\end{figure}

We have calculated $G(M)$ for the homogeneous spinless fermion model~(\ref{eq:ham}) using the procedure
introduced in the previous section. The results are shown in Fig.~\ref{fig:homogeneous}
for several interaction strengths $V$ besides the noninteracting case.
We see that $G(M)$ converges with increasing $M$ toward the exact result given by Eqs.~(\ref{eq:luttinger}) and~(\ref{eq:bethe}) 
in all cases.  

In all our figures we show polynomial fits to our data for $G(M)$. Actually,
these polynomial fits do not always yield accurate results for the extrapolation $G=\lim_{M\rightarrow \infty} G(M)$
in interacting chains. For instance, see the case $V=2t$ in Fig.~\ref{fig:homogeneous}. 
Probably, there are slowly-decaying non-analytical finite-size corrections to $G$ as a function of $1/M$.
Thus the polynomial fits should be considered as guides to the eyes.

Nevertheless, Fig.~\ref{fig:homogeneous} confirms that our method can evaluate the conductance of homogeneous Luttinger liquids.
As for the noninteracting chain, only a short wire length $M_{\text{W}}$ is required but 
the total system size $M$ (or more precisely the ratio $M/M_{\text{W}}$) must be very large to approach the thermodynamic limit quantitatively.

\subsection{Luttinger liquid with one barrier}

Field-theoretical methods~\cite{apel82a,kane92a,Kane1992} predict that impurities affect the transport properties
of Luttinger liquids in a fundamentally different way from normal metals.
Thus we apply our method to the problem of a Luttinger liquid with one and two barriers (on-site impurities) in order
to test its validity for inhomogeneous systems and verify the field-theoretical predictions in a lattice model.
We discuss first the results for a single barrier. Results for two barriers are presented in the next section.

To model the single impurity, a local potential $\epsilon$ is applied at a site $j_a$ close to the middle of the wire 
[$j_{a} \approx (j_1+j_2)/2$].
The system Hamiltonian is then
\begin{equation}
H_{\text{I}} = H + \epsilon \, n_{j_a}
\end{equation}
where $H$ is the Hamiltonian~(\ref{eq:ham}).
As $H_{\text{I}}$ is particle-hole symmetric, the conductance is independent from the sign of $\epsilon$ and thus we discuss
only the cases $\epsilon \geq 0$.

\begin{figure}
\includegraphics[width=.9\columnwidth]{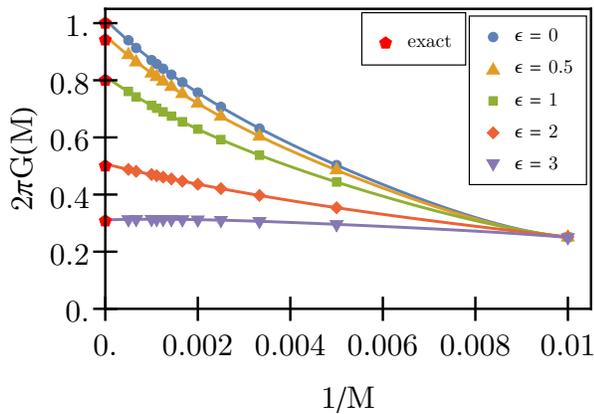}
\caption{
DMRG results for the conductance $G(M)$ of a noninteracting chain with a single on-site impurity
as a function of the inverse system size for various barrier strengths $\epsilon$. 
Solid lines are polynomial fits. 
For all cases $\eta = 48t/M$ and $M_{\text{W}}=10$. 
Pentagons on the vertical axis ($1/M=0$) show the exact values predicted by the 
Landauer formula~(\ref{eq:landauer}) and~(\ref{eq:transmission}).
}
\label{fig:landauer}
\end{figure}

First, we investigate the noninteracting chain.
According to Landauer transport theory, the local impurity can be viewed as a barrier that scatters charge carriers elastically. 
The conductance of a single-channel wire is determined by the transmission probability $T$ through the barrier at the Fermi
energy and is given by the Landauer formula~\cite{nazarov09}
\begin{equation}
G_{\text{L}} = G_0 T.
\label{eq:landauer}
\end{equation}
The transmission coefficient can easily be calculated  
for a single on-site impurity in a noninteracting one-dimensional tight-binding lattice
\begin{align}
T = & \frac{4t^2\sin^2(k_{F})}{4t^2\sin^2(k_{F})+\epsilon^2}
\label{eq:transmission}
\end{align}
where $k_{F}$ is the Fermi wave number, which takes the value $k_{F}=\pi/2$ in our half-filled model.
Thus the conductance of a noninteracting wire with a single on-site impurity is known exactly.
It decreases continuously from $G_0$ to $0$ as the barrier height $\epsilon$ increases.

We have computed the conductance of this noninteracting system using our DMRG-based method as a first test
for inhomogeneous systems. The results for $G(M)$ are shown in Fig.~\ref{fig:landauer} together with the exact results.
We see that our results match the exact values remarkably well even when the barrier height becomes large.

\begin{figure}
\includegraphics[width=.9\columnwidth]{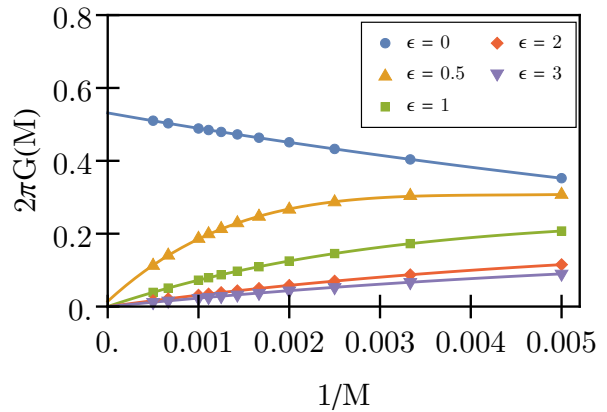}
\caption{
DMRG results for the conductance $G(M)$ of a chain with a repulsive interaction $V=2t$ and 
a single barrier of strength $0 \leq \epsilon \leq 3t$
as a function of the inverse system size.  
Solid lines are polynomial fits. 
For all cases $\eta = 48t/M$ and $M_{\text{W}}=10$. 
}
\label{fig:repulsive}
\end{figure}

\begin{figure}
\includegraphics[width=.9\columnwidth]{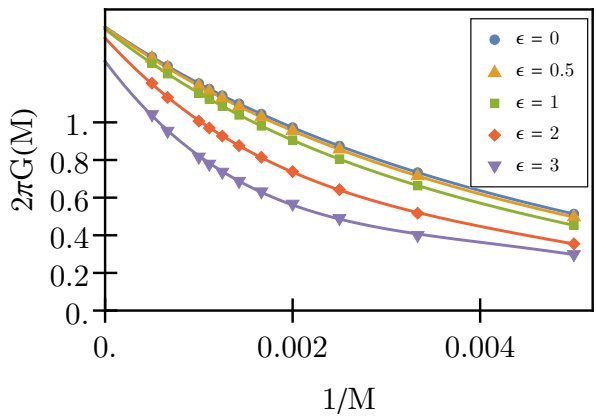}
\caption{
DMRG results for the conductance $G(M)$ of a chain with an attractive interaction $V=-t$ and 
a single barrier of strength $0 \leq \epsilon \leq 3t$
as a function of the inverse system size.  
Solid lines are polynomial fits. 
For all cases $\eta = 48t/M$ and $M_{\text{W}}=10$. 
}
\label{fig:attractive}
\end{figure}

The effects of impurities on the transport properties of Luttinger liquids are in striking
contrast to that for noninteracting fermions~\cite{kane92a,Kane1992}.
For repulsive interactions, the conductance is completely suppressed by the weakest on-site potential
and thus $G$ jumps from $G_{LL}$ to 0 as soon as $\epsilon\neq 0$.
For attractive interactions, charge carriers are not affected regardless of the barrier strength
and thus $G=G_{\text{LL}}$ for all $\epsilon$.
Our DMRG results are compatible with these universal properties as shown exemplarily in Figs.~\ref{fig:repulsive} and~\ref{fig:attractive} 
for a repulsive ($V=2t$) and an attractive ($V=-t$) chain, respectively. 

We can see for the repulsive system in Fig.~\ref{fig:repulsive} that the conductance $G(M)$ converges to zero in the thermodynamic limit
for any $\epsilon\neq 0$. In that figure, we shown again the conductance without barrier ($\epsilon=0$) from Fig.~\ref{fig:homogeneous}
to underscore the qualitatively different scaling of an homogeneous Luttinger liquid. 
Furthermore, in this enlarged scale we see more clearly that extrapolating $G(M)$ based on the polynomial fit
yields a value that slightly but visibly deviates from the exact result $G=G_{\text{LL}}=G_0/2$ 
for $\epsilon=0$, see Eqs.~(\ref{eq:luttinger}) and~(\ref{eq:bethe}),
as discussed in the previous section.

For the attractive system, we see in Fig.~\ref{fig:attractive} that $G(M)$ diminishes only slightly for increasing barrier strength
$\epsilon$ at a fixed chain length. For not too strong on-site potentials ($\epsilon \alt t$), $G(M)$ clearly
converges toward the same conductance as the homogeneous Luttinger liquid in the thermodynamic limit.
For stronger impurities, the convergence is less clear and significantly larger system sizes than $M=2000$ would be required to 
obtain a more precise extrapolation for $M\rightarrow \infty$.

In summary, these results confirm that our method can evaluate the conductance of Luttinger liquid with one impurity.
The qualitatively correct behavior is obtained for small wire lengths $M_{\text W}$.
Large system sizes $M$ are required, however, to obtain quantitative results, especially
for weak barriers (small $\epsilon$) in the repulsive case and for strong barriers (large $\epsilon$)
in the attractive case.

\subsection{Luttinger liquid with two barriers}

If the wire contains more than one barrier, the transport properties are more complicated and no longer universal,
e.g., they depend on the barrier strengths and positions.  Here we focus on the case of a repulsive Luttinger
liquid with the nearest-neighbor interaction $V=2t$.
To model two barriers in the wire, the system Hamiltonian becomes  
\begin{equation}
H_{\text{II}} = H + \epsilon_a \, n_{j_a} + \epsilon_b \, n_{j_b} 
\end{equation}
where $H$ is the Hamiltonian~(\ref{eq:ham}) and both impurity sites $j_a$ and $j_b$ are situated close to the
middle of the wire segment [$j_{a,b} \approx (j_1+j_2)/2$].
We will discuss only cases with on-site potentials $\vert  \epsilon_a \vert = \vert  \epsilon_b \vert$

Kane and Fisher~\cite{Kane1992} provided a simple physical explanation for the drastic effect
of a single barrier in a repulsive Luttinger liquid [$K<1 \Leftrightarrow V>0$ in the model~(\ref{eq:ham})].
In such a system, there is a tendency toward the formation of a charge density wave (CDW) quasi-long-range order. 
An arbitrary weak on-site potential pins the CDW, resulting in an insulating state.
In the half-filled model~(\ref{eq:ham}) the dominant CDW fluctuations have a periodicity of 2 sites. Thus
the corresponding CDW ground-state has a density profile $\langle n_j \rangle = 1/2 + (-1)^{j} \delta n$
and is twofold degenerate (i.e., $\delta n > 0$ or $\delta n <0$).
Therefore, when two barriers are added to the system, 
we have to consider two cases. First, both on-site potentials can reinforce each other, i.e., favor the same CDW ground state.
Second, both on-site potentials can oppose each other, i.e., favor different CDW ground states.

We have tested the first case for two configurations: 
(I) two next-nearest-neighbor barriers with the same potential sign ($j_a = j_b + 2, \epsilon_a =\epsilon_b$)
and (II) two nearest-neighbor barriers with potentials of opposite signs ($j_a = j_b + 1, \epsilon_a = - \epsilon_b$).
In both cases, the DMRG results for $G(M)$ are qualitatively similar to those shown in Fig.~\ref{fig:repulsive}
for a single barrier and we conclude that the conductance vanishes for any non-zero barrier height.

\begin{figure}
\includegraphics[width=.9\columnwidth]{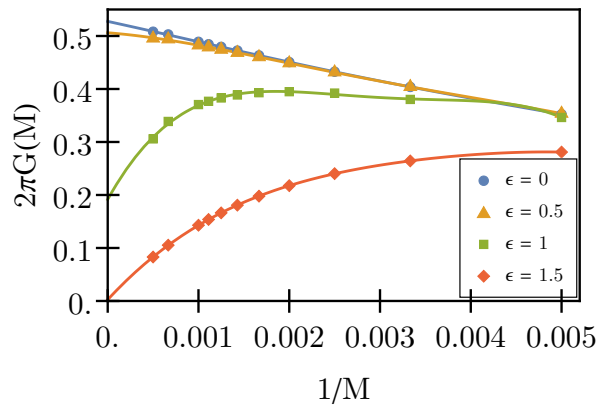}
\caption{
DMRG results for the conductance $G(M)$ of a repulsive Luttinger liquid ($V=2t$) with two non-resonant barriers
of absolute strengths $\epsilon$ as a function of the inverse system size.
The impurity sites are nearest neighbors and have the same on-site potential.  
Solid lines are polynomial fits. 
For all cases $\eta = 48t/M$ and $M_{\text{W}}=10$. 
}
\label{fig:nonuniversal}
\end{figure}

The physics is more interesting when both on-site potentials oppose each other.
As predicted by field theory~\cite{Kane1992} the transport properties are no longer universal and
can depend on the system parameters.
Figure~\ref{fig:nonuniversal} shows $G(M)$ for two nearest-neighbor barriers ($j_a = j_b + 1$)
with identical on-site potentials ($\epsilon_a = \epsilon_b = \epsilon$). We see that $G(M)$ appears to converge
to a finite value close to $G_{\text{LL}}$ for a weak barrier with $\epsilon=0.5t$ but clearly
converge to zero for a stronger barrier with $\epsilon=1.5t$. For intermediate values of $\epsilon$,
polynomial fits yield values $G_{\text{LL}}  > G > 0$ in the thermodynamic limit and thus suggest
a continuous behavior of $G$ with $\epsilon$ as in a noninteracting wire.
However, we need to investigate much larger system size $M$ to determine accurately the asymptotic 
value $G$ for intermediate barrier strengths $\epsilon$ before we can draw a conclusion.

\begin{figure}
\includegraphics[width=.9\columnwidth]{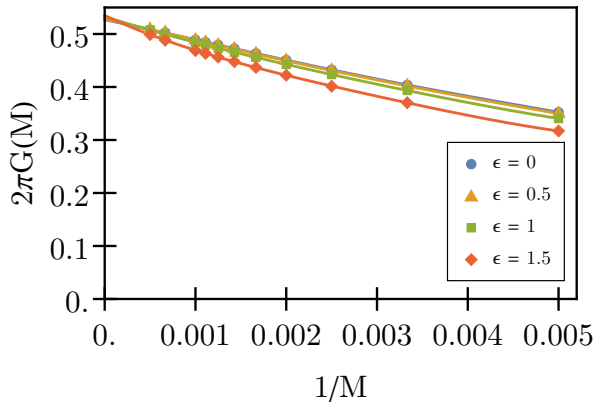}
\caption{
DMRG results for the conductance $G(M)$ of a repulsive Luttinger liquid ($V=2t$) with two resonant barriers
of absolute strengths $\epsilon$ as a function of the inverse system size.
The impurity sites are next-nearest neighbors and have on-site potentials of opposite signs. 
Solid lines are polynomial fits. 
For all cases $\eta = 48t/M$ and $M_{\text{W}}=10$. 
}
\label{fig:resonance}
\end{figure}

One of the most counterintuitive predictions of field theory~\cite{Kane1992} is that
a double barrier can exhibit a perfect resonant transmission for fine-tuned conditions
despite the fact that a single barrier causes total reflection. 
We have found that such a resonant double barrier is realized for two next-nearest-neighbor on-site
potentials of opposite signs ($j_a = j_b + 2, \epsilon_a = - \epsilon_b$).
Figure~\ref{fig:resonance} shows that $G(M)$ converges to the same value $G \approx G_{\text{LL}} = G_0/2$ 
for all tested barrier strengths. 

Therefore, our DMRG-based method is able to reproduce
some of the most striking correlation effects on the transport properties
of one-dimensional quantum systems.
As already mentioned in the other cases, only a small wire length $M_{\text{W}}$ is necessary to
find the qualitatively correct behavior, but a very
large ratio $M/M_{\text{W}} > 200$ is required to obtain accurate quantitative results for the conductance
in some unfavorable cases.

\section{Conclusion}

We have improved the DMRG calculation of the linear conductance in correlated one-dimensional lattice models
using the Kubo formula.
Our method can reproduce several properties predicted by field theory for the Luttinger liquid phase of the
half-filled spinless fermion model.
The key idea is the proper finite-size scaling, in particular for the broadening $\eta$.
A practical problem is that our approach requires a large system size, or more precisely,
a large ratio between total system size and wire length.
Nevertheless, the most difficult DMRG simulations carried out for this work (for system size $M=2000$) 
require less than 100 hours on a single modern CPU each.
Therefore, larger systems (and thus more accurate results) are certainly possible if one uses 
supercomputer facilities.

Our method can easily be extended to more general models. We have already tested it successfully on the one-dimensional
Hubbard model for interacting electrons away from half filling. In addition, it should
be possible to compute the conductance of any system for which the current-current correlation functions~(\ref{eq:correlation})
can be evaluated efficiently around $\omega=0$ using DMRG, such as electron-phonon systems~\cite{jeck07}, disordered wires, 
or ladder systems~\cite{Nocera2016,Nocera2016a}.

We tested our method on the spinless fermion model in the setup described in Sec.~\ref{sec:conductance}
because it is a well defined problem with reliable results from field theory~\cite{apel82a,kane92a,Kane1992}.
Essential features are that the wire is distinguished from the leads by the potential profile
only and that the potential difference is a model parameter.
The relevance of this setup for transport experiments is controversial, however~\cite{kawa07}.
Therefore, it is desirable, and we think that it is possible, to extend the present approach to more realistic setups
for comparison with experiments.

First, we can certainly use different interaction and hopping parameters in the wire and in the leads
to represent their different nature
and also include relatively extended and smooth transition regions.
Preliminary results confirm that our method can be applied to such systems
but also suggest that the finite-size scaling becomes more complicated.
Second, we should measure the effective potential difference between both leads in the lattice model rather than use the 
applied potential difference $V_{\text{SD}}$ to define the conductance~\cite{kawa07}.
We think that it is possible to calculate this effective potential difference from the changes
in the local density of states for a varying applied potential, which can also be calculated with dynamical DMRG~\cite{jeck13}.
Finally, it should be possible to extend our approach to finite temperatures using recently developed 
DMRG algorithms for computing frequency-resolved dynamical correlation functions at finite temperature~\cite{Tiegel2014}.
Therefore, we believe that the DMRG evaluation of the Kubo formula will become a very useful tool to study
the conductance of low-dimensional correlated systems.

\begin{acknowledgments}
J.B. thanks the Lower Saxony PhD-Programme \textit{Contacts in Nanosystems} for financial support. 
The computations presented here were partially carried out on the cluster system at the Leibniz Universit\"at Hannover, Germany.
\end{acknowledgments}

\bibliographystyle{biblev1}
\bibliography{mybibliography.bib}{}

\end{document}